\documentclass[twoside,leqno]{article}

\pdfoutput=1

\usepackage{RR}
\usepackage[a4paper]{geometry}
\usepackage{a4wide}

\usepackage[english]{babel}
\usepackage[numbers]{natbib}

\usepackage{caption}
\usepackage{fontawesome5}

\usepackage[utf8]{inputenc}
\usepackage[OT1]{fontenc} 
\usepackage{hyperref}
\usepackage{amssymb, amsmath, amsthm, mathtools, bm}
\usepackage{amsfonts, bbold, dsfont}
\usepackage{enumerate} 

\newtheorem*{definition*}{Definition}

\usepackage{listings, lstcoq}
\newcommand{\code}[1]{\texttt{#1}}

\newcommand{\iconlk}{\tiny\faExternalLink*}

\newcommand{\cnaURL}{%
  https://lipn.univ-paris13.fr/rocqdoc-num-analysis/2.1/NumAnalysis}
\newcommand{\genlk}[2]{%
  \hspace{0.1em}\raisebox{0.5ex}{\href{\cnaURL.#1.html#2}{\iconlk}}}
\bgroup\catcode`\#=12\gdef\hash{#}\egroup
  \newcommand{\lk}[2]{\genlk{#1}{\hash{#2}}}
\newcommand{\coqelk}[2]{\mbox{\coqe{#2}\lk{#1}{#2}}}
\newcommand{\filelk}[3]{\mbox{\soft{#3}\genlk{#1.#2}{}}}

\newcommand{\FLKSUB}[2]{\filelk{Subset.Subsets}{#1}{#2}}
\newcommand{\flkBR}{\FLKSUB{Binary_relation}{Binary\_relation}}
\newcommand{\FLKFF}[2]{\filelk{Subset.Subsets.FF}{#1}{#2}}
\newcommand{\flkFF}{\FLKFF{Finite_family}{Finite\_family}}
\newcommand{\flkLex}{\FLKFF{Lex}{Lex}}
\newcommand{\FLKALG}[3]{\filelk{Algebra.#1}{#2}{#3}}
\newcommand{\FLKMON}[2]{\FLKALG{Monoid}{#1}{#2}}
\newcommand{\flkMO}{\FLKMON{Monomial_order}{Monomial\_order}}
\newcommand{\FLKFEM}[2]{\filelk{FEM}{#1}{#2}}
\newcommand{\flkMI}{\FLKFEM{multi_index}{multi\_index}}

\newcommand{\LKSUB}[2]{\lk{Subset.Subsets.#1}{#2}}
\newcommand{\lkBR}[1]{\LKSUB{Binary_relation}{#1}}
\newcommand{\LKFF}[2]{\lk{Subset.Subsets.FF.#1}{#2}}

\newcommand{\lkLex}[1]{\LKFF{Lex}{#1}}
\newcommand{\LKALG}[2]{\lk{Algebra.#1}{#2}}
\newcommand{\LKMON}[2]{\LKALG{Monoid.#1}{#2}}
\newcommand{\lkMO}[1]{\LKMON{Monomial_order}{#1}}

\newcommand{\COQELKSUB}[2]{\coqelk{Subset.Subsets.#1}{#2}}
\newcommand{\coqelkBR}[1]{\COQELKSUB{Binary_relation}{#1}}
\newcommand{\COQELKFF}[2]{\coqelk{Subset.Subsets.FF.#1}{#2}}
\newcommand{\coqelkFF}[1]{\COQELKFF{Finite_family}{#1}}

\newcommand{\COQELKALG}[2]{\coqelk{Algebra.#1}{#2}}
\newcommand{\COQELKMON}[2]{\COQELKALG{Monoid.#1}{#2}}
\newcommand{\coqelkMO}[1]{\COQELKMON{Monomial_order}{#1}}

\usepackage[framemethod=tikz]{mdframed}

\usepackage{tikz}
\usepackage{circuitikz}
\usetikzlibrary{arrows.meta,calc,patterns,patterns.meta,%
  positioning,shadings,shapes}
\tikzset{math3d/.style={x= {(-0.55cm,-0.55cm)}, z={(0cm,1cm)},y={(1cm,0cm)}}}

\usepackage{color}

\definecolor{tanIII}{RGB}{205,135,63}
\definecolor{yellowII}{RGB}{238,238,0}
\definecolor{darkOliveGreenIII}{RGB}{162,205,90}
\definecolor{turquoiseII}{RGB}{0,229,238}
\definecolor{turquoiseOO}{RGB}{176,221,232}

\definecolor{darkred}{rgb}{0.7,0.2,0.2}
\definecolor{darkgreen}{rgb}{0.2,0.7,0.2}
\definecolor{darkblue}{rgb}{0.2,0.2,0.7}
\definecolor{lightred}{RGB}{255,225,225}
\definecolor{lightgreen}{RGB}{200,255,200}
\definecolor{lightblue}{RGB}{225,225,255}

\definecolor{darkgray}{gray}{0.45}
\definecolor{gray}{gray}{0.60}
\definecolor{lightgray}{gray}{0.75}
\definecolor{lightlightgray}{gray}{0.90}

\def\iscolor{1}  
\if\iscolor1

\else

\fi

\newcommand{\WP}[2]{%
  \url{https://en.wikipedia.org/w/index.php?title=#1&oldid=#2}}

\newcommand{\AuthorsList}{S. Boldo, F. Clément, V. Martin, M. Mayero}

\newcommand{\LMF}{%
  Université Paris-Saclay, Inria, CNRS, ENS Paris-Saclay,
  Laboratoire Méthodes Formelles,
  91190, Gif-sur-Yvette, France.}
\newcommand{\SERENA}{%
  a. Centre Inria de Paris, 48 rue Barrault, CS 61534,
    75647 Paris Cedex, France.\protect\\
  b. CERMICS, École des Ponts, 77455 Marne-la-Vallée, France.}
\newcommand{\LMAC}{%
  Université de technologie de Compiègne, LMAC,
  60203 Compiègne, France.}
\newcommand{\LIPN}{%
  LIPN, Université Paris 13 - USPN, CNRS UMR 7030,
  93430 Villetaneuse, France.}

\newcommand{\Funding}{%
  This work was partly supported by the European Research Council (ERC) under
  the European Union's Horizon 2020 Research and Innovation Programme – Grant
  Agreement n$^\circ$810367.
}

\newcommand{\Title}{A Rocq Formalization of Monomial and Graded Orders}
\newcommand{\TitleRR}{\Title}
\newcommand{\TitreRR}{Une formalisation en Rocq des ordres monomiaux et gradués}

\newcommand{\Abstract}{%
  Even if binary relations and orders are a common formalization topic, we need
  to formalize specific orders (namely monomial and graded) in the process of
  formalizing in Rocq the finite element method.
  This article is therefore definitions, operators, and proofs of properties
  about relations and orders, thus providing a comprehensive Rocq library.
  We especially focus on monomial orders, that are total orders compatible with
  the monoid operation.
  More than its definition and proved properties, we define several of them,
  among them the lexicographic and grevlex orders.
  For the sake of genericity, we formalize the grading of an order, a
  high-level operator that transforms a binary relation into another one, and
  we prove that grading an order preserves many of its properties, such as the
  monomial order property.
  This leads us to the definition and properties of four different graded
  orders, with very factorized proofs.
  We therefore provide a comprehensive and user-friendly library in Rocq about
  orders, including monomial and graded orders, that contains more than 700
  lemmas.
}
\newcommand{\Resume}{%
  Même si les relations binaires et les ordres sont un sujet de formalisation
  courant, nous avons besoin de formaliser des ordres spécifiques (à savoir
  monomial et gradué) dans le processus de la formalisation en Rocq de la
  méthode des éléments finis.
  Cet article traite donc des définitions, des opérateurs et des preuves des
  propriétés des relations et des ordres, fournissant ainsi une bibliothèque
  Rocq complète.
  Nous nous concentrons tout particulièrement sur les ordres monomiaux, qui
  sont des ordres totaux compatibles avec l'opération de monoïde.
  Au-delà de la définition et des propriétés prouvées, nous en définissons
  plusieurs, dont les ordres lexicographique et grevlex.
  Par souci de généricité, nous formalisons la gradation d'un ordre, un
  opérateur de haut niveau qui transforme une relation binaire en une autre, et
  nous prouvons que la gradation d'un ordre conserve bon nombre de ses
  propriétés, telle que celle d'ordre monomial.
  Cela nous mène à la définition et aux propriétés de quatre ordres gradués
  différents, avec des preuves très factorisées.
  Nous mettons donc à disposition une bibliothèque Rocq sur les ordres complète
  et facile à utiliser, incluant les ordres monomiaux et gradués, qui contient
  plus de 700 lemmes.
}
\newcommand{\Keywords}{%
  Rocq proof assistant,
  formalization of mathematics,
  formal library,
  orders
}
\newcommand{\Motscles}{%
  Assistant de preuve Rocq,
  formalisation des mathématiques,
  bibliothèque formelle,
  ordres
}

\newcommand{\eg}{e.g.}

\newcommand{\ie}{i.e.}

\newcommand{\nonstrict}{non\-strict}
\newcommand{\nontotal}{non\-total}

\newcommand{\unintuitive}{un\-intuitive}

\newcommand{\soft}[1]{\textsf{#1}}

\newcommand{\Rocq}{\soft{Rocq}}
  \newcommand{\Coquelicot}{\soft{Coquelicot}}
  
  \newcommand{\RocqNumAnalysis}{\soft{rocq-num-analysis}}

  \newcommand{\MathComp}{\soft{MathComp}}
    \newcommand{\MathCompURL}{https://math-comp.github.io/htmldoc_2_4_0}
  
  \newcommand{\SSReflect}{\soft{SSReflect}}

\newcommand{\Git}{\soft{Git}}

\newcommand{\IsabelleHOL}{\soft{Isabelle/HOL}}
  \newcommand{\IsabelleHOLURL}{https://isabelle.in.tum.de/library/HOL}
\newcommand{\Isabelle}{\soft{Isabelle}}
  \newcommand{\IsabelleAFPURL}{https://www.isa-afp.org/sessions}

\newcommand{\Lean}{\soft{Lean}}
  \newcommand{\mathlib}{\soft{mathlib}}
    \newcommand{\mathlibURL}{https://github.com/leanprover-community/mathlib4/tree/3c3f56a/Mathlib}

\newcommand{\Mizar}{\soft{Mizar}}

\newcommand{\Opam}{\soft{Opam}}

\renewcommand{\leq}{\leqslant}

\newcommand{\makespace}[1]{\quad#1\quad}

\newcommand{\Equiv}{\Longleftrightarrow}

\newcommand{\Conj}{\land}

\newcommand{\CONJ}{\makespace{\Conj}}

\newcommand{\Disj}{\lor}
\newcommand{\DISJ}{\makespace{\Disj}}

\newcommand{\eqdef}{\stackrel{\text{def.}}{=}}

\newcommand{\equivdef}{\stackrel{\text{def.}}{\Equiv}}
\newcommand{\EQUIVDEF}{\makespace{\equivdef}}

\newcommand{\N}{\mathbb{N}}

\newcommand{\Nd}{\N^d}

\newcommand{\matP}{\mathbb{P}}

\newcommand{\R}{\mathbb{R}}

\newcommand{\Rd}{\R^d}

\newcommand{\Rbar}{\overline{\R}}

\newcommand{\T}{T}
\newcommand{\Td}{\T^d}
\newcommand{\Tdmi}{\T^{d-1}}

\newcommand{\K}{K}
\newcommand{\Kd}{\K^d}

\newcommand{\matPgen}[2]{\matP^{#1}_{#2}}
\newcommand{\matPdk}{\matPgen{d}{k}}

\newcommand{\calA}{\mathcal{A}}

\newcommand{\calC}{\mathcal{C}}

\newcommand{\calS}{\mathcal{S}}

\newcommand{\calAgen}[2]{\calA^{#1}_{#2}}
\newcommand{\calAdk}{\calAgen{d}{k}}

\newcommand{\calCd}[1]{\calC^d_{#1}}
\newcommand{\calCdo}{\calCd{0}}
\newcommand{\calCdi}{\calCd{1}}
\newcommand{\calCdii}{\calCd{2}}
\newcommand{\calCdiii}{\calCd{3}}

\newcommand{\Rlt}{<}
\newcommand{\Rlto}[1]{\Rlt^{\text{#1}}}
\newcommand{\Rltlex}{\Rlto{lex}}
\newcommand{\Rltcolex}{\Rlto{colex}}
\newcommand{\Rltsymlex}{\Rlto{symlex}}
\newcommand{\Rltrevlex}{\Rlto{revlex}}

\newcommand{\Rle}{\leq}
\newcommand{\Rleo}[1]{\Rle^{\text{#1}}}
\newcommand{\Rlelex}{\Rleo{lex}}

\newcommand{\Rel}{\prec}
\newcommand{\Relo}[1]{\Rel^{\text{#1}}}
\newcommand{\Relex}{\Relo{lex}}
\newcommand{\Relcolex}{\Relo{colex}}
\newcommand{\Relsymlex}{\Relo{symlex}}
\newcommand{\Relrevlex}{\Relo{revlex}}

\newcommand{\len}[1]{\left|#1\right|}
\newcommand{\aalpha}{{\boldsymbol{\alpha}}}

\newcommand{\bbeta}{{\boldsymbol{\beta}}}

\newcommand{\ggamma}{{\boldsymbol{\gamma}}}

\renewcommand{\aa}{{\bf a}}

\newcommand{\ww}{{\bf w}}
\newcommand{\xx}{{\bf x}}

\newcommand{\XX}{{\bf X}}
\newcommand{\yy}{{\bf y}}
\newcommand{\zzero}{{\bf 0}}

\newcommand{\haa}{\hat{\aa}}

\newcommand{\caalpha}{\check{\aalpha}}
\newcommand{\cbbeta}{\check{\bbeta}}

\newcommand{\taalpha}{{\widetilde{\aalpha}}}
\newcommand{\tbbeta}{{\widetilde{\bbeta}}}

\newcommand{\baalpha}{{\overline{\aalpha}}}
\newcommand{\bbbeta}{{\overline{\bbeta}}}

\newcommand{\vect}[1]{\overrightarrow{#1}}

\newcommand{\ltb}[1]{<_{{#1}}}

\newcommand{\ltK}{\ltb{\K}}
\newcommand{\lto}[1]{<^{\text{#1}}}

\newcommand{\ltNd}{<} 
\newcommand{\ltMon}{<} 
\newcommand{\ltlex}{\lto{lex}}
\newcommand{\ltcolex}{\lto{colex}}
\newcommand{\ltrevlex}{\lto{revlex}}
\newcommand{\ltsymlex}{\lto{symlex}}
\newcommand{\ltgrlex}{\lto{grlex}}
\newcommand{\ltgrevlex}{\lto{grevlex}}
\newcommand{\ltgrcolex}{\lto{grcolex}}
\newcommand{\ltgrsymlex}{\lto{grsymlex}}

\newcommand{\st}{{\,|\,}} 

\RRdate{December 2025}

\RRauthor{%
  Sylvie Boldo%
  \thanks{{\LMF}\texttt{sylvie.boldo@inria.fr}}
  \and François Clément%
  \thanks{{\SERENA}\texttt{francois.clement@inria.fr}}
  \and Vincent Martin%
  \thanks{{\LMAC} \texttt{vincent.martin@utc.fr}}
  \and Micaela Mayero%
  \thanks{{\LIPN}\goodbreak \texttt{mayero@lipn.univ-paris13.fr}}
}
\authorhead{\AuthorsList}

\RRtitle{\TitreRR}
\RRetitle{\TitleRR}
\titlehead{\TitleRR}

\RRnote{\Funding}

\RRresume{\Resume}
\RRabstract{\Abstract}

\RRmotcle{\Motscles}
\RRkeyword{\Keywords}

\RRprojets{Toccata and Serena}

\RCSaclay

\begin{document}

\RRNo{9604}
\makeRR

\tableofcontents
\listoffigures

\section{Introduction}
\label{s:intro}

Orders are underlying fundamental objects in both mathematics and computer
science.
Moreover, they are a common formalization topic, due to their inherent
generality and usefulness.
Monomial orders are useful for tackling multivariate polynomials.
The most common applications are in commutative algebra and algorithmic
algebraic geometry.
For example, they are central to the calculation of Gröbner bases, where they
are used to define the dominant term of a polynomial, thereby facilitating the
application of Buchberger's algorithm~\cite{fro:igb:97,bw:gba:98,mar:bgo:08} or
Faugère (F4 and F5) algorithms, {\eg} see~\cite{clo:iva:15}.
Division algorithms in multivariate polynomial rings also exploit these orders
to efficiently reduce polynomials to a simplified form~\cite{ccs:stc:99}.

Some orders, such as the lexicographic order, are used for the elimination of
variables in the resolution of systems of polynomial equations or for the
projection of algebraic varieties~\cite{clo:iva:15}.
Other applications include triangulation or ideal decomposition, in connection
with combinatorial geometry and the properties of polytopes~\cite{stu:gbc:96}.
The choice of an order also influences the performance of symbolic computation
software such as Macaulay2,\footnote{\url{https://macaulay2.com/}}
Singular\footnote{\url{https://www.singular.uni-kl.de/}} or
SageMath,\footnote{\url{https://www.sagemath.org/}} directly affecting the
complexity of calculations and the structure of objects~\cite{her:bei:10}.

Monomial orders are also underlying in the field of numerical analysis for
polynomial approximations, even if it is not explicitly stated.
Our motivation is the formalization in {\Rocq} of (parts of) the Finite Element
Method, see~\cite{BCF17,BCF22,BCMMM25}, that is popular to solve partial
differential equations.
In the process, it is needed to represent $d$-multivariate polynomials of
bounded total degree $k$, often called $\matPdk$, see for
instance~\cite{BCMMM25}.
Indeed, in the Finite Element Method, it is very common to search the
approximation to a unknown function by continuous and piecewise $\matPdk$
functions.
It is thus necessary to deal with multi-indices whose sum is at most~$k$.
The following mathematical definition offers numerous possibilities for
formalization:
\begin{definition*}
  Let~$d\geq1$.
  Let~$k\in\N$.
  The {\em set of multi-indices of sum at most~$k$} is denoted~$\calAdk$, and
  is defined by
  \begin{equation*}
    \calAdk \eqdef \{ \aalpha \in \Nd \st \len{\aalpha} \leq k \},
    \quad \mbox{where } \forall \aalpha \in \Nd, \
      \len{\aalpha} \eqdef \sum_{i=0}^{d-1}\alpha_i.
  \end{equation*}
\end{definition*}
\noindent
When implementing in a program or when formalizing in \Rocq,%
\footnote{\url{https://rocq-prover.org}} one needs to order these
multi-indices, as they are lists/vectors of lists/vectors.
For more on the chosen ordering and {\Rocq} implementation of~$\calAdk$, see
Section~\ref{s:Adk-order}.

\begin{figure*}[htb]
  \centering
  \resizebox{0.8\linewidth}{!}{\begin{tikzpicture}[scale=4,math3d] 

  \def\kk{3}

  \def\colk{black}
  \def\colo{magenta}
  \def\coli{darkgreen}
  \def\colii{red}
  \def\coliii{blue}

  \def\opacity{0.4}
  \def\opacityi{0.7}
  \def\opacityii{0.6}

  \coordinate (AA) at (0,0,-0.25);
  \coordinate (CC) at ($(AA) + (0,1,0)$);
  \coordinate (DD) at ($(AA) + (0,0,1)$);
  \draw[line width=1.0pt,rounded corners=0.5pt] (AA) -- (CC) -- (DD) -- cycle;

  \fill[color=\colo] (AA) circle (0.8pt);
  \node[color=\colo,below] (Nxyz) at (AA) {$(0,0)$};
  \newcount\y
  \foreach \x in {1,0} {
    \pgfmathsetcount{\y}{1-\x} 
    \coordinate (Axyz) at ($(AA) + 1/\kk*(0,\x,\y)$);
    \node[color=\coli,below] (Nxyz) at (Axyz) {$(\x,\the\y)$};
    \fill[color=\coli] (Axyz) circle (0.8pt);
  }
  \newcount\y
  \foreach \x in {2,1,...,0} {
    \pgfmathsetcount{\y}{2-\x} 
    \coordinate (Axyz) at ($(AA) + 1/\kk*(0,\x,\y)$);
    \node[color=\colii,below] (Nxyz) at (Axyz) {$(\x,\the\y)$};
    \fill[color=\colii] (Axyz) circle (0.8pt);
  }
  \newcount\y
  \foreach \x in {3,2,...,0} {
    \pgfmathsetcount{\y}{3-\x} 
    \coordinate (Axyz) at ($(AA) + 1/\kk*(0,\x,\y)$);
    \node[color=\coliii,below] (Nxyz) at (Axyz) {$(\x,\the\y)$};
    \fill[color=\coliii] (Axyz) circle (0.8pt);
  }

  \coordinate (eps) at ($1/\kk*(0,0.05,-0.05)$); 
  \coordinate (A10) at ($(AA) + 1/\kk*(0,1,0)$);
  \coordinate (A01) at ($(AA) + 1/\kk*(0,0,1)$);
  \coordinate (A20) at ($(AA) + 1/\kk*(0,2,0)$);
  \coordinate (A02) at ($(AA) + 1/\kk*(0,0,2)$);

  \node[color=\colo,below=12pt] (Nxyz) at (AA) {$\len{\aalpha}=0$}; 
  \node[color=\coli,below=12pt] (Nxyz) at (A10) {$\len{\aalpha}=1$}; 
  \node[color=\colii,below=12pt] (Nxyz) at (A20) {$\len{\aalpha}=2$}; 
  \node[color=\coliii,below=12pt] (Nxyz) at (CC) {$\len{\aalpha}=3$}; 
  
  \draw[color=\coli,line width=0.7pt] (A10) -- ($(A01) + (eps)$);
  \draw[color=\colii,line width=0.7pt] (A20) -- ($(A02) + (eps)$);
  \draw[color=\coliii,line width=1.4pt] (CC) -- ($(DD) + (eps)$);


  \coordinate (A) at (0,2.5,0);
  \coordinate (B) at ($(A) + (1,0,0)$);
  \coordinate (C) at ($(A) + (0,1,0)$);
  \coordinate (D) at ($(A) + (0,0,1)$);
  \draw[line width=1.0pt,rounded corners=0.5pt] (A) -- (B) -- (D) -- cycle;
  \draw[line width=1.0pt,rounded corners=0.5pt] (A) -- (C) -- (B) -- cycle;
  \draw[line width=1.0pt,rounded corners=0.5pt] (B) -- (C) -- (D) -- cycle;
  \draw[line width=1.6pt,rounded corners=0.5pt] (A) -- (C) -- (D) -- cycle;


  \pgfmathparse{1-1/\kk}\let\kkp\pgfmathresult
  \coordinate (B1) at ($\kkp*(A) + 1/\kk*(B)$);
  \coordinate (C1) at ($\kkp*(A) + 1/\kk*(C)$);
  \coordinate (D1) at ($\kkp*(A) + 1/\kk*(D)$);
  \pgfmathparse{1-2/\kk}\let\kkp\pgfmathresult
  \coordinate (B2) at ($\kkp*(A) + 2/\kk*(B)$);
  \coordinate (C2) at ($\kkp*(A) + 2/\kk*(C)$);
  \coordinate (D2) at ($\kkp*(A) + 2/\kk*(D)$);

  \draw[color=\coli,fill=\coli!40,fill opacity=\opacity] (B1) -- (C1) -- (D1) -- cycle;
  \draw[color=\colii,fill=\colii!20,fill opacity=\opacity] (B2) -- (C2) -- (D2) -- cycle;
  \draw[color=\coliii,fill=\coliii!20,fill opacity=\opacity] (B) -- (C) -- (D) -- cycle;

  \fill[color=\colo,fill opacity=\opacityii] (A) circle (0.8pt);
  \foreach \x in {1,0} {
    \pgfmathparse{1-\x}\let\YY\pgfmathresult
    \foreach \y in {0,...,\YY} {
      \pgfmathparse{1-\x-\y}\let\z\pgfmathresult
      \fill[color=\coli,fill opacity=\opacityii] ($(A) + 1/\kk*(\x,\y,\z)$) circle (0.8pt);
    }
  }
  \foreach \x in {2,1,...,0} {
    \pgfmathparse{2-\x}\let\YY\pgfmathresult
    \foreach \y in {0,...,\YY} {
      \pgfmathparse{2-\x-\y}\let\z\pgfmathresult
      \fill[color=\colii,fill opacity=\opacityii] ($(A) + 1/\kk*(\x,\y,\z)$) circle (0.8pt);
    }
  }
  \newcount\z
  \foreach \x in {3,2,...,0} {
    \pgfmathparse{3-\x}\let\YY\pgfmathresult
    \foreach \y in {0,...,\YY} {
      \pgfmathsetcount{\z}{3-\x-\y} 
      \coordinate (Axyz) at ($(A) + 1/\kk*(\x,\y,\z)$);
      \node[color=\coliii,below=-0.15] (Nxyz) at (Axyz) {$(\x,\y,\the\z)$};
      \fill[color=\coliii] (Axyz) circle (0.8pt);
    }
  }


\end{tikzpicture}}
  \caption[Examples of multi-indices]{%
    Multi-indices~$\calAdk$ for~$d\in\{2,3\}$ and~$k=3$.
    Each multi-index is depicted as a colored ball.
    The colors correspond to constant sums of multi-indices.
    For instance, in blue, we depict multi-indices whose sum is~$3$.}
  \label{f:lag-k3-d2-d3}
\end{figure*}
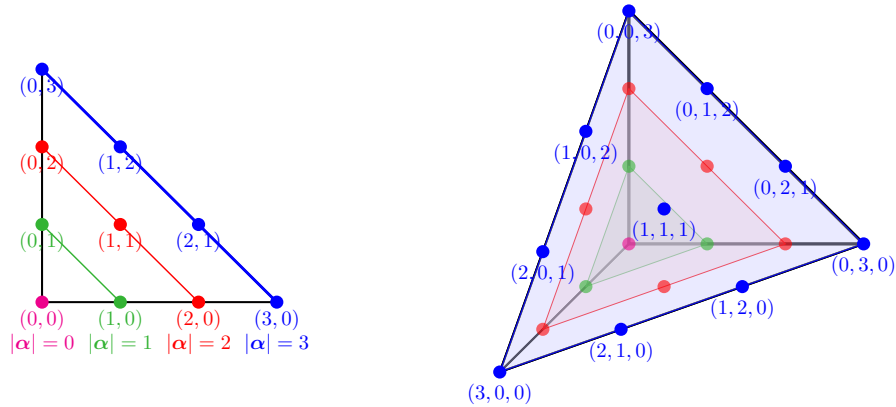

In order to illustrate an order on~$\Nd$, we use the representation of the
multi-indices on the frame $(O,x_0,\dots,x_{d-1})$ of~$\Rd$ ($d=2$ or~$3$), see
Figure~\ref{f:lag-k3-d2-d3}.
Note that the vector variables are noted with bold letters (such as~$\aalpha$).
The zero vector is denoted by~$\zzero$.

One of the objectives of this article is to provide a comprehensive and
user-friendly library that formalizes in {\Rocq} various orders on~$\Nd$ to
order the multi-index set~$\calAdk$ (or more generally on~$\mathbb{M}^d$, for
any ordered commutative monoid~$\mathbb{M}$).
Such a multi-index can be used for instance as the multi-exponent of the
$d$-multivariate monomial $\XX^\aalpha\eqdef\prod_{i=0}^{d-1}X_i^{\alpha_i}$
(whose total degree is~$\len{\aalpha}$).
Any polynomial in $\matPdk$ of total degree at most~$k$ is a linear combination
of these monomials.
Then, for instance to define a leading term, it is interesting to use an order
on monomials that is compatible with the multiplicative monoid structure of
monomials, or equivalently to use an order on multi-indices that is compatible
with the additive monoid structure of multi-indices.
Such total orders are said \emph{monomial}, and the mapping
$(\aalpha\mapsto\XX^\aalpha)$ becomes a monotonic ordered monoid morphism.
Namely, we consider a total order ``$\ltNd$'' on multi-indices that satisfies:
for all $\aalpha,\bbeta,\ggamma\in\Nd$, $\aalpha\ltNd\bbeta$ implies
$\aalpha+\ggamma\ltNd\bbeta+\ggamma$.
Or equivalently, we consider a total order ``$\ltMon$'' on monomials that
satisfies $\XX^\aalpha\ltMon\XX^\bbeta$ implies
$\XX^\aalpha\XX^\ggamma\ltMon\XX^\bbeta\XX^\ggamma$.
It is also generally required that for all nonzero~$\aalpha$,
$1=\XX^\zzero\ltMon\XX^\aalpha$, or equivalently $\zzero\ltNd\aalpha$.
Moreover, it can be of interest to have a {\em graded} monomial order, {\ie}
that respects $\len{\aalpha}<\len{\bbeta}$ (in~$\N$) implies
$\aalpha\ltNd\bbeta$ (in~$\Nd$).
Note also that monomial orders are usually defined as strict orders, but our
formalization is more generic, as it encompasses strict and {\nonstrict}
orders, with adequate definitions.

This development is freely available as part of {\Opam}%
\footnote{\url{https://rocq-prover.org/p/rocq-num-analysis/2.1.0}}
packages, the {\Git} repository is at\\
\centerline{\url{https://lipn.univ-paris13.fr/rocq-num-analysis/-/tree/2.1/}}\\
and the {\Rocq} code and the documentation are also browsable at\\
\centerline{\url{https://lipn.univ-paris13.fr/rocqdoc-num-analysis/2.1/}}\\
In the following, {\iconlk} after a definition, a lemma or a file name is the
corresponding hyperlink to the documentation.

\bigskip

Section~\ref{s:soa} is devoted to some state of the art about binary relations,
and lexicographic, monomial and graded orders.
In Section~\ref{s:bin-relations}, we briefly recall and formalize the necessary
properties on binary relations.
The Section~\ref{s:lex-orders} is dedicated to lexicographic orders, while
Section~\ref{s:mon-orders} deals with the formalization in {\Rocq} of graded
and monomial orders.
In Section~\ref{s:concl}, we detail the ordering of~$\calAdk$ and give some
conclusive remarks and perspectives.

\section{State of the Art}
\label{s:soa}

Homogeneous binary relations such as order relations are fundamental objects in
mathematics, just as functions, and are closely related to naive set theory.
Moreover, the lexicographic order, or dictionary order, is a popular way to
extend an order on items to an order on vectors (or homogeneous tuples).
Therefore, they are formalized in most proof systems.
Note that several concepts can be found under different names in the
literature, and therefore also in formalizations.
For example, preorder and quasi-order are usually synonymous, as well as total
order and linear order on the one hand, and converse, dual, transpose, flip and
inverse relation on the other hand.
We first describe, ordered by proof assistant, some formalizations on the order
properties, before going into the vector orders, including the lexicographic
and monomial ones.

In {\IsabelleHOL},\footnote{\url{\IsabelleHOLURL/}} there is support for
abstract binary relations (\soft{HOL/Relation}) and orders
(\soft{HOL/Order\_Relation}).
Relations are represented either as sets of pairs, or as binary predicates.
Most elementary properties are defined: (ir)reflexivity, (a,anti)symmetry,
transitivity, and totality.
Orders are defined as conjunctions of elementary properties: preorder, partial
order, and (strict) linear order.
Moreover, two implementations provide support for ordered structures as type
classes, either embedding both strict and {\nonstrict} versions
(\soft{HOL/Orderings}), or not (\soft{HOL-Lattice/Orders}, where lattices are
also obviously covered).
Many results are provided about the converse/dual operator (see
Section~\ref{s:bin-relations}).
In addition, the lexicographic order is defined on pairs as an operator on
relations on both item (\soft{HOL-Library/Product\_Lexorder}), with a specific
definition for the strict version and for the {\nonstrict} version.
This operator is shown to transport preorders, partial and linear orders.

In {\Lean}, the {\mathlib} library has a strong corpus on order theory,
including lattices.\footnote{\url{\mathlibURL/}}
Relations are represented as binary predicates, and elementary properties,
including trichotomous, as well as conjunctive properties, including strict
weak order, are defined as type classes
(\soft{Order/Defs/\linebreak[0]Unbundled}).
There is also specific support for ordered structures embedding both strict and
{\nonstrict} versions (\soft{Order/Defs/PartialOrder} and
\soft{Order/Defs/LinearOrder}).
The support for lexicographic order is rather versatile in \Lean.
For instance, it is provided for pairs (\soft{Data/Prod/Lex}), for lists
(\soft{Data/List/Lex}), but also for finitely supported functions
(\soft{Data/Finsupp/Lex}).
And there is also support for generic lexicographic orders on sigma-types
(\soft{Sigma/Lex}) and pi-types (\soft{Order/PiLex}), taking as arguments
relations on the indices and on each summand/item.

In \Rocq, at least three formalizations are present in the standard library.%
\footnote{\url{https://rocq-prover.org/doc/V8.20.0/stdlib/}}
(i) The \soft{Sets} standard library provides basic support for some of the
usual definitions; it is quite old-fashioned.
(ii) The \soft{Relations} standard library also provides some support for
elementary properties such as reflexivity, (anti)symmetry, and transitivity.
Then, (pre)order is defined as a record collecting the appropriate elementary
properties.
An inductive definition provides the lexicographic order on pairs.
This formalization is used in several parts of the standard library such as
\soft{Sorting}, \soft{Wellfounded} and \soft{micromega} (for \soft{Lra}).
(iii) The \soft{Classes} standard library uses the type \coqe{relation} from
\soft{Relations}, then redefines elementary and conjunctive properties as type
classes, including irreflexivity and asymmetry, strict and partial orders.
There are lemmas for interactions of the flip operator with all elementary and
conjunctive properties.
This formalization is used to implement the generalized rewrite tactic for
setoids.
It is also used in several parts of the standard library such as \soft{Logic},
\soft{MSets}, \soft{NArith}, \soft{PArith} and \soft{Reals}.

The main definitions are also provided in the {\SSReflect} part of the standard
library.
This is extended in the {\MathComp} library~\cite{MathComp_Ref},%
\footnote{\url{\MathCompURL/mathcomp.ssreflect.order.html}}
which provides the usual aspects of order theory with full support for
types equipped with partial or total orders, and lattices.
Each algebraic structure is equipped with both the {\nonstrict} and strict
variants of the order.

\bigskip

Monomial orders (also called admissible orders or term orders) and graded
orders (also called degree orders) can be found as a part of formal
developments about the computation of Gröbner bases in polynomial rings.
Note that the well-order property can be included in the definition of monomial
order, in particular when the termination of multivariate polynomial division
algorithms is at stake.

In the {\Isabelle} Archive of Formal Proofs,%
\footnote{\url{\IsabelleAFPURL/polynomials/\#Power_Products.html}}
lexicographic orders are defined from any well-founded order and degree orders
are defined from any scalar linear order and any relation on vectors.
Then, degree-lexicographic orders and degree-reverse-lexicographic orders are
defined.

In \Lean, the {\mathlib} library provides support for monomial orders that
embed the well-founded property (\soft{Data/Finsupp/MonomialOrder}).
Then, the degree lexicographic order is defined as the lexicographic order
applied to pairs formed by the sum of the vector and the vector itself
(\soft{Data/Finsupp/MonomialOrder/DegLex}).

In \Mizar, Schwarzweller~\cite{Sch06} uses an abstract term order for the
reduction of polynomials and the building of Gröbner bases.

In \Rocq, we can cite the work by Perez~\cite{Perez2004} in which Gröbner bases
needed some developments on polynomial rings using a generic order relation.
We can also cite the works by Théry~\cite{The01} and \soft{grobner}.%
\footnote{\url{https://github.com/thery/grobner/}}
The former is based on the \soft{Relations} standard library.
The main order definition combines a well-founded preorder with compatibility
with some monoid law; it is instantiated with the multiplication of monomials.
A lexicographic order is defined for monomials only (or equivalently on tuples
of natural numbers).
The latter is based on the {\MathComp} library.

\bigskip

For comparison, we do not cover lattices, or well-founded orders, but we
provide many definitions including uncommon ones such as the union and
intersection operators for binary relations, the connected, trichotomous and
negatively transitive elementary properties, strict total order and strict weak
order.
We also provide many equivalent definitions, and a fairly comprehensive
exploration of interactions between operators, elementary properties and
conjunctive properties, resulting in about 300 lemmas (see
Section~\ref{s:bin-relations}).
We place particular emphasis on providing definitions and properties that are
close to standard mathematics.
Furthermore, as this work is part of our {\RocqNumAnalysis} library, we cannot,
for example, use {\MathComp} canonical structures that are not compatible with
those of {\Coquelicot} and our library.

An original feature is our definition for the lexicographic order that unifies
the strict and {\nonstrict} cases, while taking care of the empty tuple.
And the definition for graded orders is not restricted to the sole graded
lexicographic order, \coqe{graded} is a generic operator that takes any scalar
relation to compare sums, and any vector relation to then compare vectors.
Again, we provide a fairly comprehensive corpus of lemmas about the various
flavors of lexicographic orders, graded orders, monomial orders, and their
interactions with basic operators and properties.
This results in nearly 400 additional lemmas (see Sections~\ref{s:lex-orders}
and~\ref{s:mon-orders}).

\section{Basic Operators and Properties of Binary Relations}
\label{s:bin-relations}

Before dealing with monomial and graded orders, we first need to define the
basics of order theory with some operators and properties about binary
relations.
The simplest example that comes to mind is the usual order on natural numbers,
but this can be source of misleading intuition since that order actually
possess many more properties than a mere (partial) order.
For instance, it is a total order, and such that its complementary is also a
total order.
Instead, it can be preferable to take the inclusion in the power set of some
set as mental image.
It is not total, and its complementary is not an order since it is not
transitive.
We used~\cite{ehr:mo:05} as reference, but we also took some inspiration from
the related Wikipedia pages.%
\footnote{\WP{Binary_relation}{1299977865}\\
  \WP{Homogeneous_relation}{1289819519}}

In the file {\flkBR}, given any type~\coqe{T} and a relation
\coqe{R : T -> T -> Prop}, we first define operators taking~\coqe{R} and
outputting a relation such as
\begin{lstlisting}
Definition converse?\lkBR{converse}? : T -> T -> Prop := \fun x y => R y x.
Definition complementary?\lkBR{complementary}? : T -> T -> Prop := \fun x y => ~ R x y.
Definition br_or_eq?\lkBR{br_or_eq}? : T -> T -> Prop := \fun x y => x = y \/ R x y.
\end{lstlisting}

Then, we define the following elementary properties that are used for the
various flavors of orders
\begin{lstlisting}
Definition transitive?\lkBR{transitive}? : Prop := \forall x y z, R x y -> R y z -> R x z.
Definition negatively_transitive?\lkBR{negatively_transitive}? : Prop := \forall x y z, ~ R x y -> ~ R y z -> ~ R x z.
Definition reflexive?\lkBR{reflexive}? : Prop := \forall x, R x x.
Definition irreflexive?\lkBR{irreflexive}? : Prop := \forall x, ~ R x x.
Definition antisymmetric?\lkBR{antisymmetric}? : Prop := \forall x y, R x y -> R y x -> x = y.
Definition asymmetric?\lkBR{asymmetric}? : Prop := \forall x y, R x y -> ~ R y x.
Definition connected?\lkBR{connected}? : Prop := \forall x y, x <> y -> R x y \/ R y x.
Definition strongly_connected?\lkBR{strongly_connected}? : Prop := \forall x y, R x y \/ R y x.
Definition trichotomous?\lkBR{trichotomous}? : Prop :=
  \forall x y, (x = y /\ ~ R x y /\ ~ R y x) \/ (x <> y /\ R x y /\ ~ R y x) \/ (x <> y /\ R y x /\ ~ R x y).
\end{lstlisting}
Note that connected and strongly connected respectively mean that the
complementary is antisymmetric and asymmetric.
Finally, we define conjunctive properties such as
\begin{lstlisting}
Definition total_order?\lkBR{total_order}? : Prop :=
  transitive R /\ reflexive R /\ antisymmetric R /\ strongly_connected R /\
  negatively_transitive R /\ connected R.
Definition strict_total_order?\lkBR{strict_total_order}? : Prop :=
  transitive R /\ irreflexive R /\ asymmetric R /\ connected R /\
  negatively_transitive R /\ antisymmetric R /\ trichotomous R.
Definition strict_weak_order?\lkBR{strict_weak_order}? : Prop :=
  negatively_transitive R /\ irreflexive R /\ asymmetric R /\
  transitive R /\ antisymmetric R.
\end{lstlisting}
The latter actually means that the complementary is a total preorder ({\ie}
lacking the antisymmetry property).
Note that these definitions are stated for usability in the form of
conjunctions of relevant elementary properties.
Then, equivalences with less hypotheses are provided as lemmas for shorter
proofs, such as
\begin{lstlisting}
Lemma sto_correct?\lkBR{sto_correct}? :
  strict_total_order R <-> transitive R /\ irreflexive R /\ asymmetric R /\ connected R.
Lemma sto_equiv_no_asym?\lkBR{sto_equiv_no_asym}? :
  strict_total_order R <-> transitive R /\ irreflexive R /\ connected R.
Lemma sto_equiv_no_irrefl?\lkBR{sto_equiv_no_irrefl}? :
  strict_total_order R <-> transitive R /\ asymmetric R /\ connected R.
\end{lstlisting}

Besides equivalent definitions, we also provide a fairly comprehensive corpus
of lemmas on interactions between operators, elementary properties, and
conjunctive properties.
For instance, the \coqe{converse} operator is compatible with all properties,
such as
\begin{lstlisting}
Lemma conv_asym?\lkBR{conv_asym}? : asymmetric R -> asymmetric (converse R).
Lemma conv_to?\lkBR{conv_to}? : total_order R -> total_order (converse R).
\end{lstlisting}
And since \coqe{converse} is involutive, equivalences also hold.
We also strived to provide the minimal hypotheses to obtain some specific
property from another, or some equivalence.
For instance, we have
\begin{lstlisting}
Lemma asym_equiv_irrefl_antisym?\lkBR{asym_equiv_irrefl_antisym}? : asymmetric R <-> irreflexive R /\ antisymmetric R.
Lemma irrefl_asym_equiv?\lkBR{irrefl_asym_equiv}? : transitive R -> irreflexive R <-> asymmetric R.
Lemma tricho_equiv_asym_conn?\lkBR{tricho_equiv_asym_conn}? : trichotomous R <-> asymmetric R /\ connected R.
Lemma conn_str_conn?\lkBR{conn_str_conn}? : reflexive R -> connected R -> strongly_connected R.
Lemma trans_neg_trans?\lkBR{trans_neg_trans}? : connected R -> transitive R -> negatively_transitive R.
Lemma sto_equiv_swo?\lkBR{sto_equiv_swo}? : strict_total_order R <-> strict_weak_order R /\ connected R.
\end{lstlisting}

\section{Lexicographic Orders}
\label{s:lex-orders}

From a scalar order on a set~$\T$, we want to build a vector order on~$\Td$,
while retaining the properties of the initial scalar order.
The notation~$\Td$ refers to homogeneous tuples of size~$d$, formalized as
finite families (see below).
When~$\T$ has enough properties or by abuse, elements of~$\Td$ are also called
vectors.

We first recall the well-known lexicographic order on~$\Td$, and some others
that can be derived easily.
These are proved to be monomial orders under conditions in
Section~\ref{s:mon-orders}.

In the sequel, $\aalpha$ and~$\bbeta$ denote any finite families in~$\Td$.
We use the following notations when $d\geq2$: the check notation
$\caalpha\in\Tdmi$ denotes the {\em last} $d-1$ components of~$\aalpha$, and
the tilde notation $\taalpha\in\Tdmi$ denotes the {\em first} ones.
Thus, when $\aalpha\eqdef(\alpha_0,\ldots,\alpha_{d-1})$, we have
$\aalpha=(\alpha_0,\caalpha)=(\taalpha,\alpha_{d-1})$.
We also note $\baalpha\eqdef(\alpha_{d-1},\ldots,\alpha_0)$.
The {\Rocq} development can be found in the file {\flkLex}.

\subsection{Lexicographic Order}
\label{s:lex}

In the case of a strict order, we have the following definition of the usual
lexicographic order,
\begin{equation}
  \label{e:lex_lt}
  \aalpha \Rltlex \bbeta \EQUIVDEF
  \left\{
    \begin{array}{l}
      \alpha_0 \Rlt \beta_0, \mbox { or}\\
      \alpha_0 = \beta_0 \CONJ d \geq 2 \CONJ \caalpha \Rltlex \cbbeta.
    \end{array}
  \right.
\end{equation}
We have $\aalpha\Rltlex\bbeta$ iff $\alpha_i\Rlt\beta_i$ for the {\em first}
index~$i$ for which~$\alpha_i$ and~$\beta_i$ differ.
See a 2D~example with multi-indices ($\T=\N$) in
Figure~\ref{f:tet-tria_k3_lex_colex}.
And in the {\nonstrict} case, a simple mathematical definition can be
\begin{equation}
  \label{e:lex_le}
  \aalpha \Rlelex \bbeta \EQUIVDEF
  \aalpha = \bbeta \DISJ \aalpha \Rltlex \bbeta.
\end{equation}

\bigskip

Neither Equation~\eqref{e:lex_lt} nor~\eqref{e:lex_le} is the chosen definition
in {\Rocq} as we have preferred to combine~$\Rltlex$ and~$\Rlelex$ into a
single definition, and this is~\eqref{e:lex} we specify in {\Rocq}.
Given a scalar order~$\Rel$ on~$T$, the {\em lexicographic order}, or simply
{\em ``lex'' order}, noted~$\Relex$, can be recursively defined by
\begin{equation}
  \label{e:lex}
  \aalpha \Relex \bbeta \EQUIVDEF
  \left\{
    \begin{array}{l}
      \alpha_0 \neq \beta_0 \CONJ \alpha_0 \Rel \beta_0, \mbox { or}\\
      \alpha_0 = \beta_0 \CONJ d \geq 2 \CONJ \caalpha \Relex \cbbeta.
    \end{array}
  \right.
\end{equation}

In \Rocq, following Section~\ref{s:bin-relations}, we consider any
type~\coqe{T} (defined as implicit \coqe{Context} in {\Rocq}) and any binary
relation~\coqe{R : T -> T -> Prop} (defined as \coqe{Variable}), which does not
need to be an order.
The type of $n$-families on~\coqe{T} is \coqe{[0..n) -> T}, denoted~\coqe{'T^n}
(see file {\flkFF}), where \coqe{[0..n)} denotes the finite type
\coqe{ordinal n} from {\MathComp}~\cite{MathComp_Ref}.
Then, the application of the lex operator is of type
\coqe{lex R : \forall n, 'T^n -> 'T^n -> Prop}.
The recursive definition (provided in file {\flkLex}) uses a \coqe{Fixpoint}.
It is rather tedious, and instead, we present here the two specification lemmas
that describe the base case and the recursive case and are more readable.

The base case, where~\coqe{'T^0} is the unit type of empty families, is
specified by
\begin{lstlisting}
Lemma lex_nil?\lkLex{lex_nil}? : \forall (x y : 'T^0), lex x y <-> reflexive R
\end{lstlisting}
This means that for equal family arguments, the result of \coqe{lex R x x}
depends on whether or not~\coqe{R} is reflexive.
With this generalization, we can deal with empty arguments, and with both
strict and {\nonstrict} orders for~\coqe{R}, while keeping the same property
for \coqe{lex R}.

The recursive case (for nonempty families) is specified by
\begin{lstlisting}
Lemma lex_S?\lkLex{lex_S}? : \forall {n} {x y : 'T^n.+1}, lex x y <->
  (x ord0 <> y ord0 /\ R (x ord0) (y ord0)) \/ (x ord0 = y ord0 /\ lex R (skipF0 x) (skipF0 y)).
\end{lstlisting}
Note that the type \coqe{'T^n.+1} (where the notation \coqe{n.+1} stands for
the successor of~\coqe{n}) structurally grants that the family arguments are
nonempty.
There are two possibilities, exactly corresponding to~\eqref{e:lex}:
\begin{itemize}
\item either the first items~$x_0$ and~$y_0$ differ (in \Rocq, \coqe{x ord0}
  and \coqe{y ord0}), and then they must be related by~\coqe{R};
\item or the first items are equal, and then \coqe{lex R} is recursively called
  on the remaining of the families where the first item is skipped (in \Rocq,
  using \coqelkFF{skipF0}).
\end{itemize}

Then, for example, to use lex with the strict and nonstrict orders on natural
numbers, we simply define the notations \coqe{lex_lt := (lex lt)} and
\coqe{lex_le := (lex le)}.
And, we prove the connection between~\eqref{e:lex_lt} and~\eqref{e:lex_le} in
this case,
\begin{lstlisting}
Lemma lex_le_lt?\lkLex{lex_le_lt}? : \forall {n}, @@lex _ le n = br_or_eq (lex lt).
\end{lstlisting}

\bigskip

Then, three variants of this \coqe{lex} operator can be defined by either
reverting the order of the items of the family arguments (colex order), or by
swapping the arguments (symlex order), or both (revlex order).
As mentioned in the introduction, the choice for a specific order depends on
the goal pursued.

\subsection{Colexicographic Order}
\label{s:colex}

The {\em colexicographic order}, or {\em ``colex'' order}, is defined by
\begin{equation}
  \label{e:colex}
  \aalpha \Relcolex \bbeta \EQUIVDEF \baalpha \Relex \bbbeta.
\end{equation}
It is similar to the lex order, but starting from the right.
For a strict order, we have $\aalpha\Rltcolex\bbeta$ iff $\alpha_i\Rlt\beta_i$
for the {\em last} index~$i$ for which~$\alpha_i$ and~$\beta_i$ differ.
This amounts to start from the left on~$\baalpha$ and~$\bbbeta$.
It is also called {\em inverse lexicographic order}, or {\em ``invlex'' order},
{\eg} see~\cite[p.~61]{clo:iva:15}.

In \Rocq, we define the colex operator as
\begin{lstlisting}
Definition colex?\lkLex{colex}? {n} : 'T^n -> 'T^n -> Prop := reverse lex.
\end{lstlisting}
where, given any vector relation \coqe{Rn : 'T^n -> 'T^n -> Prop},
\begin{lstlisting}
Definition reverse?\lkLex{reverse}? (x y : 'T^n) : Prop := Rn (revF x) (revF y).
\end{lstlisting}
and \coqelkFF{revF}\ $\aalpha:=\baalpha$.

\subsection{Symmetrical Lexicographic Order}
\label{s:symlex}

We design the {\em symmetrical lexicographic order}, or simply
{\em ``symlex'' order}, defined by
\begin{equation}
  \label{e:symlex}
  \aalpha \Relsymlex \bbeta \EQUIVDEF \bbeta \Relex \aalpha.
\end{equation}
It is the symmetrical of the lex order.
For a strict order, we have $\aalpha \Rltsymlex\bbeta$ iff
$\beta_i\Rlt\alpha_i$ for the {\em first} index~$i$ for which~$\alpha_i$
and~$\beta_i$ differ.
We have found no mention of this variant in the literature.

In \Rocq, we define the symlex operator as
\begin{lstlisting}
Definition symlex?\lkLex{symlex}? {n} : 'T^n -> 'T^n -> Prop := converse lex.
\end{lstlisting}
where \coqelkBR{converse} is the swapping of the arguments of binary relations
(see Section~\ref{s:bin-relations}).

\subsection{Reverse Lexicographic Order}
\label{s:revlex}

The {\em reverse lexicographic order}, or {\em ``revlex'' order}, is defined by
\begin{equation}
  \label{e:revlex}
  \aalpha \Relrevlex \bbeta \EQUIVDEF \bbeta \Relcolex \aalpha.
\end{equation}
It is the symmetrical of the colex order.
For a strict order, we have $\aalpha\Rltrevlex\bbeta$ iff $\beta_i\Rlt\alpha_i$
for the {\em last} index~$i$ for which~$\alpha_i$ and~$\beta_i$ differ.
It is also called {\em reverse inverse lexicographic order}, or
{\em ``rinvlex'' order}, {\eg} see~\cite[p.~61]{clo:iva:15}.

In \Rocq, we define the revlex operator as
\begin{lstlisting}
Definition revlex?\lkLex{revlex}? {n} : 'T^n -> 'T^n -> Prop := converse colex.
\end{lstlisting}

\bigskip

In Section~\ref{s:mon-orders}, the four lex-like operators are proved to
transport the monomial order property of their scalar relation argument.
Note also that lex and colex are obviously equivalent when~$d=1$ (and so are
symlex and revlex).

Additionally, we also have
$\baalpha\Relsymlex\bbbeta\Equiv\aalpha\Relrevlex\bbeta$ and
$\baalpha\Relrevlex\bbbeta\Equiv\aalpha\Relsymlex\bbeta$.
All these properties are proved in \Rocq.

\bigskip

\begin{figure*}[ht]
  \centering
  \hfill
  \resizebox{0.24\linewidth}{!}{\begin{tikzpicture}[scale=4,math3d] 

  \def\kk{3}

  \def\colk{black}
  \def\colb{blue}

  \coordinate (AA) at (0,0,-0.25);
  \coordinate (CC) at ($(AA) + (0,1,0)$);
  \coordinate (DD) at ($(AA) + (0,0,1)$);
  \draw[line width=1.0pt,rounded corners=0.5pt] (AA) -- (CC) -- (DD) -- cycle;
  \node[color=\colk] (K2) at ($(AA) + (0,0.65,0.7)$) {lex};

  \fill[color=\colk] (AA) circle (0.8pt);
  \node[color=\colk,below] (Nxyz) at (AA) {$(0,0)$};
  \newcount\y
  \foreach \x in {1,0} {
    \pgfmathsetcount{\y}{1-\x} 
    \coordinate (Axyz) at ($(AA) + 1/\kk*(0,\x,\y)$);
    \node[color=\colk,below] (Nxyz) at (Axyz) {$(\x,\the\y)$};
    \fill[color=\colk] (Axyz) circle (0.8pt);
  }
  \newcount\y
  \foreach \x in {2,1,...,0} {
    \pgfmathsetcount{\y}{2-\x} 
    \coordinate (Axyz) at ($(AA) + 1/\kk*(0,\x,\y)$);
    \node[color=\colk,below] (Nxyz) at (Axyz) {$(\x,\the\y)$};
    \fill[color=\colk] (Axyz) circle (0.8pt);
  }
  \newcount\y
  \foreach \x in {3,2,...,0} {
    \pgfmathsetcount{\y}{3-\x} 
    \coordinate (Axyz) at ($(AA) + 1/\kk*(0,\x,\y)$);
    \node[color=\colb,below] (Nxyz) at (Axyz) {$(\x,\the\y)$};
    \fill[color=\colb] (Axyz) circle (0.8pt);
  }

  \coordinate (eps0) at ($1/\kk*(0,0,-0.05)$);
  \coordinate (eps) at ($1/\kk*(0,0.05,-0.05)$);

  \coordinate (A00) at ($(AA)$);
  \coordinate (A03) at ($(AA) + 1/\kk*(0,0,3)$);
  \coordinate (A10) at ($(AA) + 1/\kk*(0,1,0)$);
  \coordinate (A12) at ($(AA) + 1/\kk*(0,1,2)$);
  \coordinate (A20) at ($(AA) + 1/\kk*(0,2,0)$);
  \coordinate (A21) at ($(AA) + 1/\kk*(0,2,1)$);
  \coordinate (A30) at ($(AA) + 1/\kk*(0,3,0)$);
  \draw[color=\colb,dashed,line width=1.4pt,->,>=latex]
    (A00) -- ($(A03) + (eps0)$);
  \draw[color=\colb,dashed,line width=1pt,->,>=latex]
    (A03) -- ($(A10) - (eps)$);
  \draw[color=\colb,dashed,line width=1pt,->,>=latex]
    (A10) -- ($(A12) + (eps0)$);
  \draw[color=\colb,dashed,line width=1pt,->,>=latex]
    (A12) -- ($(A20) - (eps)$);
  \draw[color=\colb,dashed,line width=1pt,->,>=latex]
    (A20) -- ($(A21) + (eps0)$);
  \draw[color=\colb,dashed,line width=1.4pt,->,>=latex]
    (A21) -- ($(A30) - (eps)$);

\end{tikzpicture}}
  \hfill
  \resizebox{0.24\linewidth}{!}{\begin{tikzpicture}[scale=4,math3d] 

  \def\kk{3}

  \def\colk{black}
  \def\colb{blue}

  \coordinate (AA) at (0,0,-0.25);
  \coordinate (CC) at ($(AA) + (0,1,0)$);
  \coordinate (DD) at ($(AA) + (0,0,1)$);
  \draw[line width=1.0pt,rounded corners=0.5pt] (AA) -- (CC) -- (DD) -- cycle;
  \node[color=\colk] (K2) at ($(AA) + (0,0.65,0.7)$) {colex};

  \fill[color=\colk] (AA) circle (0.8pt);
  \node[color=\colk,below] (Nxyz) at (AA) {$(0,0)$};
  \newcount\y
  \foreach \x in {1,0} {
    \pgfmathsetcount{\y}{1-\x} 
    \coordinate (Axyz) at ($(AA) + 1/\kk*(0,\x,\y)$);
    \node[color=\colk,below] (Nxyz) at (Axyz) {$(\x,\the\y)$};
    \fill[color=\colk] (Axyz) circle (0.8pt);
  }
  \newcount\y
  \foreach \x in {2,1,...,0} {
    \pgfmathsetcount{\y}{2-\x} 
    \coordinate (Axyz) at ($(AA) + 1/\kk*(0,\x,\y)$);
    \node[color=\colk,below] (Nxyz) at (Axyz) {$(\x,\the\y)$};
    \fill[color=\colk] (Axyz) circle (0.8pt);
  }
  \newcount\y
  \foreach \x in {3,2,...,0} {
    \pgfmathsetcount{\y}{3-\x} 
    \coordinate (Axyz) at ($(AA) + 1/\kk*(0,\x,\y)$);
    \node[color=\colb,below] (Nxyz) at (Axyz) {$(\x,\the\y)$};
    \fill[color=\colb] (Axyz) circle (0.8pt);
  }

  \coordinate (eps0) at ($1/\kk*(0,-0.05,0)$);
  \coordinate (eps) at ($1/\kk*(0,-0.05,0.05)$);

  \coordinate (A00) at ($(AA)$);
  \coordinate (A30) at ($(AA) + 1/\kk*(0,3,0)$);
  \coordinate (A01) at ($(AA) + 1/\kk*(0,0,1)$);
  \coordinate (A21) at ($(AA) + 1/\kk*(0,2,1)$);
  \coordinate (A02) at ($(AA) + 1/\kk*(0,0,2)$);
  \coordinate (A12) at ($(AA) + 1/\kk*(0,1,2)$);
  \coordinate (A03) at ($(AA) + 1/\kk*(0,0,3)$);
  \draw[color=\colb,dashed,line width=1.4pt,->,>=latex]
    (A00) -- ($(A30) + (eps0)$);
  \draw[color=\colb,dashed,line width=1pt,->,>=latex]
    (A30) -- ($(A01) - (eps)$);
  \draw[color=\colb,dashed,line width=1pt,->,>=latex]
    (A01) -- ($(A21) + (eps0)$);
  \draw[color=\colb,dashed,line width=1pt,->,>=latex]
    (A21) -- ($(A02) - (eps)$);
  \draw[color=\colb,dashed,line width=1pt,->,>=latex]
    (A02) -- ($(A12) + (eps0)$);
  \draw[color=\colb,dashed,line width=1.4pt,->,>=latex]
    (A12) -- ($(A03) - (eps)$);

\end{tikzpicture}}
  \hfill
  \resizebox{0.24\linewidth}{!}{\begin{tikzpicture}[scale=4,math3d] 

  \def\kk{3}

  \def\colk{black}
  \def\colb{blue}

  \coordinate (AA) at (0,0,-0.25);
  \coordinate (CC) at ($(AA) + (0,1,0)$);
  \coordinate (DD) at ($(AA) + (0,0,1)$);
  \draw[line width=1.0pt,rounded corners=0.5pt] (AA) -- (CC) -- (DD) -- cycle;
  \node[color=\colk] (K2) at ($(AA) + (0,0.65,0.7)$) {symlex};

  \fill[color=\colk] (AA) circle (0.8pt);
  \node[color=\colk,below] (Nxyz) at (AA) {$(0,0)$};
  \newcount\y
  \foreach \x in {1,0} {
    \pgfmathsetcount{\y}{1-\x} 
    \coordinate (Axyz) at ($(AA) + 1/\kk*(0,\x,\y)$);
    \node[color=\colk,below] (Nxyz) at (Axyz) {$(\x,\the\y)$};
    \fill[color=\colk] (Axyz) circle (0.8pt);
  }
  \newcount\y
  \foreach \x in {2,1,...,0} {
    \pgfmathsetcount{\y}{2-\x} 
    \coordinate (Axyz) at ($(AA) + 1/\kk*(0,\x,\y)$);
    \node[color=\colk,below] (Nxyz) at (Axyz) {$(\x,\the\y)$};
    \fill[color=\colk] (Axyz) circle (0.8pt);
  }
  \newcount\y
  \foreach \x in {3,2,...,0} {
    \pgfmathsetcount{\y}{3-\x} 
    \coordinate (Axyz) at ($(AA) + 1/\kk*(0,\x,\y)$);
    \node[color=\colb,below] (Nxyz) at (Axyz) {$(\x,\the\y)$};
    \fill[color=\colb] (Axyz) circle (0.8pt);
  }

  \coordinate (eps0) at ($1/\kk*(0,0,-0.05)$);
  \coordinate (eps1) at ($1/\kk*(0,0.05,0)$);
  \coordinate (eps2) at ($1/\kk*(0,0.05,-0.05)$);
  \coordinate (eps) at ($1/\kk*(0,-0.025,0.05)$);

  \coordinate (A00) at ($(AA)$);
  \coordinate (A03) at ($(AA) + 1/\kk*(0,0,3)$);
  \coordinate (A10) at ($(AA) + 1/\kk*(0,1,0)$);
  \coordinate (A12) at ($(AA) + 1/\kk*(0,1,2)$);
  \coordinate (A20) at ($(AA) + 1/\kk*(0,2,0)$);
  \coordinate (A21) at ($(AA) + 1/\kk*(0,2,1)$);
  \coordinate (A30) at ($(AA) + 1/\kk*(0,3,0)$);
  \draw[color=\colb,dashed,line width=1.4pt,<-,>=latex]
    ($(A00) - (eps0)$) -- (A03);
  \draw[color=\colb,dashed,line width=1pt,<-,>=latex]
    ($(A03) - (eps)$) -- (A10);
  \draw[color=\colb,dashed,line width=1pt,<-,>=latex]
    ($(A10) - (eps0)$) -- (A12);
  \draw[color=\colb,dashed,line width=1pt,<-,>=latex]
    ($(A12) - (eps)$) -- (A20);
  \draw[color=\colb,dashed,line width=1pt,<-,>=latex]
    ($(A20) - (eps0)$) -- (A21);
  \draw[color=\colb,dashed,line width=1.4pt,<-,>=latex]
    ($(A21) + (eps2)$) -- (A30);

\end{tikzpicture}}
  \hfill
  \resizebox{0.24\linewidth}{!}{\begin{tikzpicture}[scale=4,math3d] 

  \def\kk{3}

  \def\colk{black}
  \def\colb{blue}

  \coordinate (AA) at (0,0,-0.25);
  \coordinate (CC) at ($(AA) + (0,1,0)$);
  \coordinate (DD) at ($(AA) + (0,0,1)$);
  \draw[line width=1.0pt,rounded corners=0.5pt] (AA) -- (CC) -- (DD) -- cycle;
  \node[color=\colk] (K2) at ($(AA) + (0,0.65,0.7)$) {revlex};

  \fill[color=\colk] (AA) circle (0.8pt);
  \node[color=\colk,below] (Nxyz) at (AA) {$(0,0)$};
  \newcount\y
  \foreach \x in {1,0} {
    \pgfmathsetcount{\y}{1-\x} 
    \coordinate (Axyz) at ($(AA) + 1/\kk*(0,\x,\y)$);
    \node[color=\colk,below] (Nxyz) at (Axyz) {$(\x,\the\y)$};
    \fill[color=\colk] (Axyz) circle (0.8pt);
  }
  \newcount\y
  \foreach \x in {2,1,...,0} {
    \pgfmathsetcount{\y}{2-\x} 
    \coordinate (Axyz) at ($(AA) + 1/\kk*(0,\x,\y)$);
    \node[color=\colk,below] (Nxyz) at (Axyz) {$(\x,\the\y)$};
    \fill[color=\colk] (Axyz) circle (0.8pt);
  }
  \newcount\y
  \foreach \x in {3,2,...,0} {
    \pgfmathsetcount{\y}{3-\x} 
    \coordinate (Axyz) at ($(AA) + 1/\kk*(0,\x,\y)$);
    \node[color=\colb,below] (Nxyz) at (Axyz) {$(\x,\the\y)$};
    \fill[color=\colb] (Axyz) circle (0.8pt);
  }

  \coordinate (eps0) at ($1/\kk*(0,-0.05,0)$);
  \coordinate (eps) at ($1/\kk*(0,-0.05,0.05)$);

  \coordinate (A00) at ($(AA)$);
  \coordinate (A30) at ($(AA) + 1/\kk*(0,3,0)$);
  \coordinate (A01) at ($(AA) + 1/\kk*(0,0,1)$);
  \coordinate (A21) at ($(AA) + 1/\kk*(0,2,1)$);
  \coordinate (A02) at ($(AA) + 1/\kk*(0,0,2)$);
  \coordinate (A12) at ($(AA) + 1/\kk*(0,1,2)$);
  \coordinate (A03) at ($(AA) + 1/\kk*(0,0,3)$);
  \draw[color=\colb,dashed,line width=1.4pt,->,>=latex]
    (A03) -- ($(A12) + (eps)$);
  \draw[color=\colb,dashed,line width=1pt,->,>=latex]
    (A12) -- ($(A02) - (eps0)$);
  \draw[color=\colb,dashed,line width=1pt,->,>=latex]
    (A02) -- ($(A21) + (eps)$);
  \draw[color=\colb,dashed,line width=1pt,->,>=latex]
    (A21) -- ($(A01) - (eps0)$);
  \draw[color=\colb,dashed,line width=1pt,->,>=latex]
    (A01) -- ($(A30) + (eps)$);
  \draw[color=\colb,dashed,line width=1.4pt,->,>=latex]
    (A30) -- ($(A00) - (eps0)$);
\end{tikzpicture}}
  \hfill
  \caption[Lex, colex, symlex, and revlex orderings in 2D]{%
    Lex, colex, symlex, and revlex  orderings (from left to right)
    of~$\calAdk\subset\Nd$ when $d=2$ and $k=3$.
    The increase in the order is represented by dashed arrows.
    For~$\calA_3^2$, we have\\
    \centerline{%
      \begin{tabular}{ll}
        lex & $(0,0)<(0,1)<(0,2)<(0,3)<(1,0)<(1,1)<(1,2)<(2,0)<(2,1)<(3,0)$\\
        colex & $(0,0)<(1,0)<(2,0)<(3,0)<(0,1)<(1,1)<(2,1)<(0,2)<(1,2)<(0,3)$
      \end{tabular}}
    The symlex order is the symmetrical of the lex order, and the revlex order
    is the symmetrical of the colex order.
    In view of graded orders, note that when the sum of multi-indices is~3
    (hypotenuse of the triangles, blue nodes), we have
    $(0,3)\ltlex(1,2)\ltlex(2,1)\ltlex(3,0)$, and also
    $(0,3)\ltrevlex(1,2)\ltrevlex(2,1)\ltrevlex(3,0)$.}
  \label{f:tet-tria_k3_lex_colex}
\end{figure*}
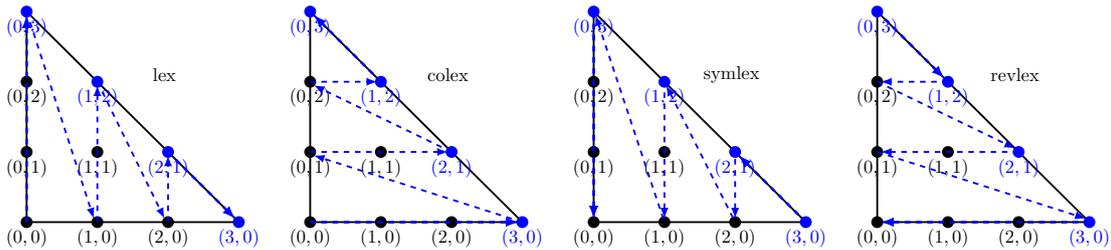

The lex order and its variants are not very convenient in practice for
multinomial ordering, as they do not sort the monomials according to their
total degree.
For instance, for $d=2$, let $p\eqdef X_0^0X_1^8$ and $q\eqdef X_0^1X_0^2$.
We have $p\ltlex q$ (as $0<1$), but $\deg(p)=8>3=\deg(q)$.
Thus, it can be useful to introduce \emph{graded} monomial orders.

\section{Monomial and Graded Orders}
\label{s:mon-orders}

Now let us consider our basic type~\coqe{G} to be an Abelian monoid, meaning we
have an identity element denoted by~0 and an operator denoted by~$+$ that is
both associative and commutative.
It corresponds to the \coqe{AbelianMonoid} canonical structure of
Coquelicot~\cite{BLM15}.
On this type, we define what a monomial order is in Section~\ref{s:mon-def},
and define what a graded order is, and its properties, in
Section~\ref{s:graded}.
The {\Rocq} development can be found in the file {\flkMO}.

\subsection{Monomial Order Definition}
\label{s:mon-def}

Given an Abelian monoid~\coqe{G}, a relation~\coqe{R} is compatible with~$+$ on
the right when
\begin{lstlisting}
Definition br_plus_compat_r?\lkMO{br_plus_compat_r}? : Prop := \forall x x1 x2, R x1 x2 -> R (x1 + x) (x2 + x).
\end{lstlisting}
Then, a monomial order is both a strict total order and
\coqe{br_plus_compat_r}.
We have a similar definition for {\nonstrict} monomial orders.
For instance, the standard strict and {\nonstrict} orders on~$\N$ are obviously
monomial (see \coqe!Nat.add_{lt,! \coqe!le}_mono_r!), even if there is no total
subtraction.
As explained in Section~\ref{s:intro}, this compatibility property formalizes
the expected assumptions on orders on monomials. Indeed, we expect that
$\XX^\aalpha<\XX^\bbeta$ implies $\XX^{\aalpha+\ggamma}=\XX^\aalpha \XX^\ggamma<
\XX^\bbeta \XX^\ggamma=\XX^{\bbeta+\ggamma}$.

We then relate it to the previous definitions and operators.
For instance, a monomial {\nonstrict} order is antisymmetric
(\coqelkMO{mons_antisym}) and strongly connected (\coqelkMO{mons_str_conn}).
We also prove that \coqe{converse R} is a monomial order iff~\coqe{R} is one,
and how it relates to the complementary.
We also have definitions and lemmas on monomial orders where zero is comparable
on the left to all nonzero elements (such as on~$\N$).

We prove that the lexicographic orders defined in Section~\ref{s:lex-orders}
(lex, colex, symlex, and revlex) are monomial orders, provided the initial
scalar order is a monomial order and the addition is regular on the right (so
that we may simplify).
\begin{lstlisting}
Lemma lex_mo?\lkMO{lex_mo}? : plus_is_reg_r G ->
  \forall {R : G -> G -> Prop}, monomial_order R -> \forall (n : nat), monomial_order (lex R).
\end{lstlisting}
There is a simplification above for readability (as \coqe{lex R} cannot guess
its~\coqe{n} implicit parameter).

\subsection{Graded Orders}
\label{s:graded}

From the monomial orders (lex, colex, symlex, and revlex) based on the
lexicographic order, see Section~\ref{s:lex-orders} and
Figure~\ref{f:tet-tria_k3_lex_colex}, it is possible to generate graded
monomial orders: grlex, grcolex, grsymlex, grevlex, see below.
The principle is to compare first the sum of the families (that is to say the
sum of the items of the vectors), with a given scalar relation on~\coqe{G}, and
then in case of equality use another given vector relation on \coqe{'G^n}.
These four orders are commonly defined recursively, after the consideration of
the first index (grlex and grsymlex), or of the last index (grcolex and
grevlex), and then, in the case of equality, the treatment of the rest of the
family.
Some of these orders are well-known (such as grevlex~\cite[p.~58]{clo:iva:15}),
other less common (grsymlex for instance).

These four orders are different when $d\geq3$.
Table~\ref{t:ex_graded_monomials} illustrates the differences between these
four graded monomial orders on a few monomials of the same total degree.

\begin{table}[h]
\begin{center}
\begin{tabular}{|l|c|c|c|c|c|c|}
  \hline
  grlex $<$  & $Z^3$    & $Y^3$ & $XYZ$ & $XY^2$ & $X^3$ \\
  \hline
  grcolex $<$ &  $X^3$    & $XY^2$ & $Y^3$ & $XYZ$ & $Z^3$ \\
  \hline
  grsymlex $<$ &  $X^3$   & $XY^2$ & $XYZ$ &  $Y^3$ & $Z^3$ \\
  \hline
  grevlex $<$ &  $Z^3$    & $XYZ$ & $Y^3$ & $XY^2$ & $X^3$ \\
  \hline
\end{tabular}
\end{center}
\caption{Ordering example on monomials with the strict order on~$\N$
  for the four graded orders presented
  here, with a constant total degree (=3).
  Increase from left to right.}
  \label{t:ex_graded_monomials}
\end{table}

\subsubsection{Mathematical Definitions of Graded Orders}

For the sake of readability, the mathematical definitions are given assuming
the scalar order is strict.

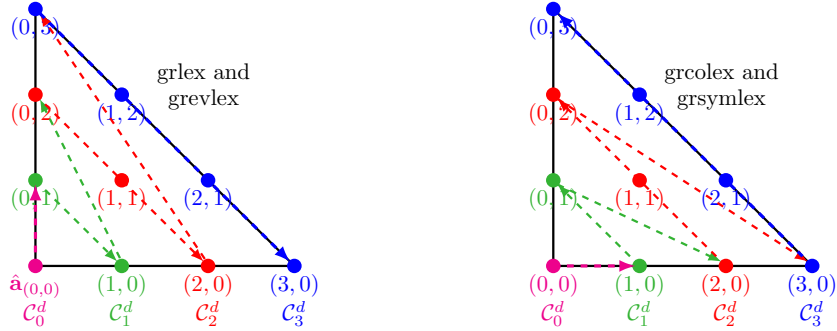
\begin{figure*}[ht]
  \centering
  \resizebox{0.75\linewidth}{!}{\begin{tikzpicture}[scale=4,math3d] 

  \def\kk{3}

  \def\colk{black}
  \def\colo{magenta}
  \def\coli{darkgreen}
  \def\colii{red}
  \def\coliii{blue}

  \def\opacity{0.4}
  \def\opacityi{0.7}
  \def\opacityii{0.6}

  \coordinate (AA) at (0,0,0);
  \coordinate (CC) at ($(AA) + (0,1,0)$);
  \coordinate (DD) at ($(AA) + (0,0,1)$);
  \draw[line width=1.0pt,rounded corners=0.5pt] (AA) -- (CC) -- (DD) -- cycle;
  \node[color=\colk] (K2) at ($(AA) + (0,0.65,0.75)$) {grlex and};
  \node[color=\colk] (K2p) at ($(AA) + (0,0.65,0.65)$) {grevlex};

  \fill[color=\colo] (AA) circle (0.8pt);
  \node[color=\colo,below] (Nxyz) at (AA) {$\haa_{(0,0)}$};
  \newcount\y
  \foreach \x in {1,0} {
    \pgfmathsetcount{\y}{1-\x} 
    \coordinate (Axyz) at ($(AA) + 1/\kk*(0,\x,\y)$);
    \node[color=\coli,below] (Nxyz) at (Axyz) {$(\x,\the\y)$};
    \fill[color=\coli] (Axyz) circle (0.8pt);
  }
  \newcount\y
  \foreach \x in {2,1,...,0} {
    \pgfmathsetcount{\y}{2-\x} 
    \coordinate (Axyz) at ($(AA) + 1/\kk*(0,\x,\y)$);
    \node[color=\colii,below] (Nxyz) at (Axyz) {$(\x,\the\y)$};
    \fill[color=\colii] (Axyz) circle (0.8pt);
  }
  \newcount\y
  \foreach \x in {3,2,...,0} {
    \pgfmathsetcount{\y}{3-\x} 
    \coordinate (Axyz) at ($(AA) + 1/\kk*(0,\x,\y)$);
    \node[color=\coliii,below] (Nxyz) at (Axyz) {$(\x,\the\y)$};
    \fill[color=\coliii] (Axyz) circle (0.8pt);
  }

  \coordinate (eps0) at ($1/\kk*(0,0,-0.05)$); 
  \coordinate (eps) at ($1/\kk*(0,0.05,-0.05)$); 
  \node[color=\colo,below=12pt] (Nxyz) at (AA) {$\calCdo$};
  \coordinate (A10) at ($(AA) + 1/\kk*(0,1,0)$);
  \coordinate (A01) at ($(AA) + 1/\kk*(0,0,1)$);
  \draw[color=\colo,dashed,line width=1.4pt,->,>=latex] (AA) -- ($(A01) + (eps0)$);
  \draw[color=\coli,dashed,line width=1pt,->,>=latex] (A01) -- ($(A10) - (eps)$);
  \node[color=\coli,below=12pt] (Nxyz) at (A10) {$\calCdi$};
  \coordinate (A20) at ($(AA) + 1/\kk*(0,2,0)$);
  \coordinate (A02) at ($(AA) + 1/\kk*(0,0,2)$);
  \draw[color=\coli,dashed,line width=1pt,->,>=latex] ($(A10) + (eps)$) -- ($(A02) + (eps)$);
  \draw[color=\colii,dashed,line width=1pt,->,>=latex] (A02) -- ($(A20) - (eps)$);
  \node[color=\colii,below=12pt] (Nxyz) at (A20) {$\calCdii$};
  \coordinate (A30) at ($(AA) + 1/\kk*(0,3,0)$);
  \coordinate (CD) at ($1/2*(CC) + 1/2*(DD)$);
  \draw[color=\colii,dashed,line width=1pt,->,>=latex] ($(A20) + (eps)$) -- ($(DD) + (eps)$);
  \draw[color=\coliii,dashed,line width=1.4pt,->,>=latex]  ($(DD) + (eps)$) -- ($(A30) - (eps)$);
  \node[color=\coliii,below=12pt] (Nxyz) at (A30) {$\calCdiii$};

  \coordinate (AA) at (0,2,0);
  \coordinate (CC) at ($(AA) + (0,1,0)$);
  \coordinate (DD) at ($(AA) + (0,0,1)$);
  \draw[line width=1.0pt,rounded corners=0.5pt] (AA) -- (CC) -- (DD) -- cycle;
  \node[color=\colk] (K2) at ($(AA) + (0,0.65,0.75)$) {grcolex and};
  \node[color=\colk] (K2p) at ($(AA) + (0,0.65,0.65)$) {grsymlex};

  \fill[color=\colo] (AA) circle (0.8pt);
  \node[color=\colo,below] (Nxyz) at (AA) {$(0,0)$};
  \newcount\y
  \foreach \x in {1,0} {
    \pgfmathsetcount{\y}{1-\x} 
    \coordinate (Axyz) at ($(AA) + 1/\kk*(0,\x,\y)$);
    \node[color=\coli,below] (Nxyz) at (Axyz) {$(\x,\the\y)$};
    \fill[color=\coli] (Axyz) circle (0.8pt);
  }
  \newcount\y
  \foreach \x in {2,1,...,0} {
    \pgfmathsetcount{\y}{2-\x} 
    \coordinate (Axyz) at ($(AA) + 1/\kk*(0,\x,\y)$);
    \node[color=\colii,below] (Nxyz) at (Axyz) {$(\x,\the\y)$};
    \fill[color=\colii] (Axyz) circle (0.8pt);
  }
  \newcount\y
  \foreach \x in {3,2,...,0} {
    \pgfmathsetcount{\y}{3-\x} 
    \coordinate (Axyz) at ($(AA) + 1/\kk*(0,\x,\y)$);
    \node[color=\coliii,below] (Nxyz) at (Axyz) {$(\x,\the\y)$};
    \fill[color=\coliii] (Axyz) circle (0.8pt);
  }

  \coordinate (eps0) at ($1/\kk*(0,-0.05,0)$); 
  \coordinate (eps) at ($1/\kk*(0,0.05,-0.05)$); 
  \node[color=\colo,below=12pt] (Nxyz) at (AA) {$\calCdo$};
  \coordinate (A10) at ($(AA) + 1/\kk*(0,1,0)$);
  \coordinate (A01) at ($(AA) + 1/\kk*(0,0,1)$);
  \draw[color=\colo,dashed,line width=1.4pt,->,>=latex] (AA) -- ($(A10) + (eps0)$);
  \draw[color=\coli,dashed,line width=1pt,->,>=latex] (A10) -- ($(A01) + (eps)$);
  \node[color=\coli,below=12pt] (Nxyz) at (A10) {$\calCdi$};
  \coordinate (A20) at ($(AA) + 1/\kk*(0,2,0)$);
  \coordinate (A02) at ($(AA) + 1/\kk*(0,0,2)$);
  \draw[color=\coli,dashed,line width=1pt,->,>=latex] ($(A01) + (eps)$) -- ($(A20) - (eps)$);
  \draw[color=\colii,dashed,line width=1pt,->,>=latex] (A20) -- ($(A02) + (eps)$);
  \node[color=\colii,below=12pt] (Nxyz) at (A20) {$\calCdii$};
  \coordinate (A30) at ($(AA) + 1/\kk*(0,3,0)$);
  \coordinate (CD) at ($1/2*(CC) + 1/2*(DD)$);
  \draw[color=\colii,dashed,line width=1pt,->,>=latex] ($(A02) + (eps)$) -- ($(A30) - (eps)$);
  \draw[color=\coliii,dashed,line width=1.4pt,->,>=latex] (A30) -- ($(DD) + (eps)$);
  \node[color=\coliii,below=12pt] (Nxyz) at (A30) {$\calCdiii$};

\end{tikzpicture}}
  \caption[Grlex, grcolex, grsymlex, and grevlex orderings in 2D]{%
    Grlex, grcolex, grsymlex, and grevlex orderings of~$\calAdk$ when $d=2$ and
    $k=3$.
    The increase in the order is represented by dashed arrows.
    For~$\calA_3^2$, grlex and grevlex are equivalent, and so are grcolex and
    grsymlex.
    We have\\
    \centerline{%
      \begin{tabular}{ll}
        grlex/grevlex
        & $(0,0)<(0,1)<(1,0)<(0,2)<(1,1)<(2,0)<(0,3)<(1,2)<(2,1)<(3,0)$\\
        grcolex/grsymlex
        & $(0,0)<(1,0)<(0,1)<(2,0)<(1,1)<(0,2)<(3,0)<(2,1)<(1,2)<(0,3)$
      \end{tabular}}}
  \label{f:tria_k3_graded}
\end{figure*}

The {\em graded lexicographic order}, or simply {\em ``grlex'' order}, is
defined by
\begin{equation*}
  \aalpha \ltgrlex \bbeta \EQUIVDEF
  \left\{
    \begin{array}{l}
      \len{\aalpha} < \len{\bbeta}, \mbox { or}\\
      \len{\aalpha} = \len{\bbeta} \CONJ \aalpha \ltlex \bbeta.
    \end{array}
  \right.
\end{equation*}
This amounts to first compare the sum of families, and in case of equality, use
the standard lex order~\eqref{e:lex}.
Thus, when $\len{\aalpha}=\len{\bbeta}$, we have $\aalpha\ltgrlex\bbeta$ iff
$\alpha_i<\beta_i$ for the first index~$i$ for which~$\alpha_i$ and~$\beta_i$
differ.
See~2D and~3D examples in Figures~\ref{f:tria_k3_graded}
and~\ref{f:tet_k3_graded}.
It is also called {\em degree lexicographic order}, or {\em ``deglex'' order}.

By inlining~\eqref{e:lex_lt} (we are in the strict case), we actually obtain
the following equivalence, which may be seen as an alternative recursive
definition,
\begin{equation*}
  \aalpha \ltgrlex \bbeta \Equiv
  \left\{
    \begin{array}{l}
      \len{\aalpha} < \len{\bbeta}, \mbox { or}\\
      \len{\aalpha} = \len{\bbeta}
      \CONJ \alpha_0 < \beta_0, \mbox{ or}\\
      \len{\aalpha} = \len{\bbeta} \CONJ \alpha_0 = \beta_0 \  \CONJ
          d \geq 2 \CONJ \caalpha \ltgrlex \cbbeta.
    \end{array}
  \right.
\end{equation*}

\begin{figure*}[ht]
  \centering
  \resizebox{0.85\linewidth}{!}{\input{fig_Tet_k3_graded}}
  \caption[Grlex, grcolex, grsymlex, and grevlex orderings in 3D]{%
    Grlex, grcolex, grsymlex, and grevlex orderings of~$\calAdk$ when
    $d=k=3$.
    The increase in the order is represented by dashed arrows, only when the
    sum $l=3$ (see Figure~\ref{f:lag-k3-d2-d3}).
    The four orders are different.
    For the multi-indices of sum~3, we have\\
    \centerline{%
    \resizebox{\textwidth}{!}{%
      \begin{tabular}{ll}
        grlex
        & $(0,0,3)<(0,1,2)<(0,2,1)<(0,3,0)<(1,0,2)<(1,1,1)<(1,2,0)<(2,0,1)<
          (2,1,0)<(3,0,0)$\\
        grcolex
        & $(3,0,0)<(2,1,0)<(1,2,0)<(0,3,0)<(2,0,1)<(1,1,1)<(0,2,1)<(1,0,2)<
          (0,1,2)<(0,0,3)$\\
        grsymlex
        & $(3,0,0)<(2,1,0)<(2,0,1)<(1,2,0)<(1,1,1)<(1,0,2)<(0,3,0)<(0,2,1)<
          (0,1,2)<(0,0,3)$\\
        grevlex
        & $(0,0,3)<(0,1,2)<(1,0,2)<(0,2,1)<(1,1,1)<(2,0,1)<(0,3,0)<(1,2,0)<
          (2,1,0)<(3,0,0)$
      \end{tabular}}}
    The restriction of grsymlex to the last two components when the sum is 3,
    is exactly grsymlex for $d=2$ (see Figure~\ref{f:tria_k3_graded}, right).}
  \label{f:tet_k3_graded}
\end{figure*}

\bigskip

We design the {\em graded colexicographic order}, or simply
{\em ``grcolex'' order}, defined by
\begin{equation*}
  \aalpha \ltgrcolex \bbeta \EQUIVDEF
  \left\{
    \begin{array}{l}
      \len{\aalpha} < \len{\bbeta}, \mbox { or}\\
      \len{\aalpha} = \len{\bbeta} \CONJ \aalpha \ltcolex \bbeta.
    \end{array}
  \right.
\end{equation*}
This amounts to first compare the sum of families, and in case of equality, use
the colex order~\eqref{e:colex}.
Thus, when $\len{\aalpha}=\len{\bbeta}$, we have $\aalpha\ltgrcolex\bbeta$ iff
$\alpha_i<\beta_i$ for the last index~$i$ where~$\alpha_i$ and~$\beta_i$
differ.
As before, we have an alternative recursive definition.
Compare on the same~2D and~3D examples the grcolex and grlex orders in
Figure~\ref{f:tria_k3_graded}.

\bigskip

We also design the {\em graded symmetric lexicographic order}, or simply
{\em ``grsymlex'' order}, defined by
\begin{equation}
  \label{e:grsymlex}
  \aalpha \ltgrsymlex \bbeta \EQUIVDEF
  \left\{
    \begin{array}{l}
      \len{\aalpha} < \len{\bbeta}, \mbox { or}\\
      \len{\aalpha} = \len{\bbeta} \CONJ \aalpha \ltsymlex \bbeta.
    \end{array}
  \right.
\end{equation}
This amounts to first compare the sum of families, and in case of equality, use
the symlex order~\eqref{e:symlex}.
Thus, when $\len{\aalpha}=\len{\bbeta}$, we have $\aalpha\ltgrsymlex\bbeta$ iff
$\beta_i<\alpha_i$ for the first index~$i$ where~$\alpha_i$ and~$\beta_i$
differ.

We have an alternative recursive definition,
\begin{equation}
  \label{e:grsymlex-equiv-1}
  \aalpha \ltgrsymlex \bbeta \Equiv
  \left\{
    \begin{array}{l}
      \len{\aalpha} < \len{\bbeta}, \mbox { or}\\
      \len{\aalpha} = \len{\bbeta}
      \CONJ \beta_0 < \alpha_0, \mbox{ or}\\
      \len{\aalpha} = \len{\bbeta} \CONJ \beta_0 = \alpha_0 \  \CONJ
         d \geq 2 \CONJ \caalpha \ltgrsymlex \cbbeta,
    \end{array}
  \right.
\end{equation}
but it may be simplified,
\begin{equation}
  \label{e:grsymlex-equiv-2}
  \aalpha \ltgrsymlex \bbeta \Equiv
  \left\{
    \begin{array}{l}
      \len{\aalpha} < \len{\bbeta}, \mbox { or}\\
      \len{\aalpha} = \len{\bbeta} \CONJ
          d \geq 2 \CONJ \caalpha \ltgrsymlex \cbbeta,
    \end{array}
  \right.
\end{equation}
as when $\len{\aalpha}=\len{\bbeta}$ and $\beta_0<\alpha_0$, we have~$d\geq2$
and $\len{\caalpha}<\len{\cbbeta}$, {\ie} $\caalpha\ltgrsymlex\cbbeta$.
Note that this simplification is made possible by the symmetric aspect of
symlex.

More precisely, grsymlex and grevlex (see below) have three equivalent
definitions (by grading, recursive, and simplified recursive) while grlex and
grcolex have only two: indeed, for instance for grlex,
$\len{\aalpha}=\len{\bbeta}\Conj\alpha_0<\beta_0$ implies
$\len{\cbbeta}<\len{\caalpha}$, {\ie} $\cbbeta\ltgrlex\caalpha$, but not
$\caalpha\ltgrlex\cbbeta$.

Note that the grcolex and grsymlex orders are identical when $d=1$ or~2, but
differ as soon as $d\geq3$.
Compare on the same~2D and~3D examples in Figure~\ref{f:tria_k3_graded}.

\bigskip

The {\em graded reverse lexicographic order}, or simply
{\em ``grevlex'' order}, is defined by
\begin{equation*}
  \aalpha \ltgrevlex \bbeta \EQUIVDEF
  \left\{
    \begin{array}{l}
      \len{\aalpha} < \len{\bbeta}, \mbox { or}\\
      \len{\aalpha} = \len{\bbeta} \CONJ \aalpha \ltrevlex \bbeta.
    \end{array}
  \right.
\end{equation*}
This amounts to first compare the sum of families, and in case of equality, use
the revlex order~\eqref{e:revlex}.
Thus, when $\len{\aalpha}=\len{\bbeta}$, we have $\aalpha\ltgrevlex\bbeta$ iff
$\beta_i<\alpha_i$ for the last index~$i$ where~$\alpha_i$ and~$\beta_i$
differ.
It is also called {\em degree reverse lexicographic order}, or simply
{\em ``degrevlex'' order}.
As before, we have a simplified recursive definition,
\begin{equation*}
  \aalpha \ltgrevlex \bbeta \Equiv
  \left\{
    \begin{array}{l}
      \len{\aalpha} < \len{\bbeta}, \mbox { or}\\
      \len{\aalpha} = \len{\bbeta}
      \CONJ d \geq 2 \CONJ \taalpha \ltgrevlex \tbbeta,
    \end{array}
  \right.
\end{equation*}
as when $\len{\aalpha}=\len{\bbeta}$ and $\beta_{d-1}<\alpha_{d-1}$
(with~$d\geq2$), we have $\len{\taalpha}<\len{\tbbeta}$, {\ie}
$\taalpha\ltgrevlex\tbbeta$.

Note that the grlex and grevlex orders are identical when $d=1$ or~2, but
differ as soon as $d\geq3$.
Compare on the same~2D and~3D examples in Figure~\ref{f:tria_k3_graded}.

\subsubsection{Formal Definitions of Graded Orders}

All theses definitions share the same pattern.
We have therefore formalized how to grade any binary relation.
More precisely, given an Abelian monoid~\coqe{G}, a scalar relation~\coqe{R}
on~\coqe{G} \emph{and} a vector relation~\coqe{Rn} on~\coqe{'G^n}, we define
another vector relation on~\coqe{'G^n},
\begin{lstlisting}
Definition graded?\lkMO{graded}? (x y : 'G^n) : Prop :=
  (sum x <> sum y /\ R (sum x) (sum y)) \/ (sum x = sum y /\ Rn x y).
\end{lstlisting}
Note that, similarly to the lexicographical order definition of
Section~\ref{s:lex}, this definition encompasses both strict and {\nonstrict}
cases for~\coqe{R}.

The previous definitions are then straightforward,
\begin{lstlisting}
Definition grlex?\lkMO{grlex}? := graded R (lex R).
Definition grcolex?\lkMO{grcolex}? := graded R (colex R).
Definition grsymlex?\lkMO{grsymlex}? := graded R (symlex R).
Definition grevlex?\lkMO{grevlex}? := graded R (revlex R).
\end{lstlisting}
We also provide all the various recursive definitions as equivalences, with
some hypotheses on~\coqe{G} and~\coqe{R} when needed (such as
\coqelkMO{grsymlex_S} for~\eqref{e:grsymlex-equiv-1} and
\coqelkMO{grsymlex_S_mo} for~\eqref{e:grsymlex-equiv-2}).

More interesting, many properties hold about the \coqe{graded} operator, making
a large factorization of proofs on all the wanted lemmas on the graded
lexicographic orders.
For instance,
the reflexivity or irreflexivity of the relation \coqe{graded R Rn} is that
of~\coqe{Rn} ({\eg} see \coqelkMO{graded_irrefl_equiv}),
while the symmetry, antisymmetry and asymmetry needs the ones of both~\coqe{R}
and~\coqe{Rn} ({\eg} see \coqelkMO{graded_asym}).
A comprehensive study of the properties of Section~\ref{s:bin-relations} is
formalized.

To get back to monomial orders, we prove that the graded of monomial orders is
a monomial order,
\begin{lstlisting}
Lemma graded_mo?\lkMO{graded_mo}? : plus_is_reg_r G ->
  monomial_order R -> monomial_order Rn -> monomial_order (graded R Rn).
\end{lstlisting}
It is therefore straightforward to prove that \coqe{grsymlex lt} is a monomial
order (\coqelkMO{grsymlex_lt_mo}).

An unexpected lemma is about the idempotence of the \coqe{graded} operator with
respect to its scalar relation argument,
\begin{lstlisting}
Lemma graded_idem?\lkMO{graded_idem}? : \forall {R1} R2 {n} {Rn : 'G^n -> 'G^n -> Prop},
  graded R1 (graded R2 Rn) = graded R1 Rn.
\end{lstlisting}
When grading with respect to~\coqe{R2} (on~\coqe{G}) and then~\coqe{R1}
(on~\coqe{G} too), it is equivalent to only grading with respect to~\coqe{R1}.
This may allow us to remove one graded out of two.
For instance, when \coqe{R1=R2}, it means that
\coqe{graded R (grsymlex R) = grsymlex R}.
And for different orders on~$\N$, we can prove for instance that
\coqe{graded gt (grcolex lt) = graded gt (colex lt)}.

\section{Application, Conclusion \& Perspectives}
\label{s:concl}

\subsection{Application: Ordering of $\calAdk$}
\label{s:Adk-order}

As explained in Section~\ref{s:intro}, we are interested in defining and
ordering $\calAdk\eqdef\{\aalpha\in\Nd\st\len{\aalpha}\leq k\}$, that
corresponds to the exponents of $\matPdk$, the $d$-multivariate polynomials of
degree smaller or equal to~$k$ (see file {\flkMI}).
As we want the corresponding family to be ordered by degree, we rely on a
graded order and we construct~$\calAdk$ by slices of increasing degrees,
\[
  \calAdk = \biguplus_{l=0}^k  \calC^d_l,
  \qquad \mbox{with }
  \calC^d_l \eqdef \{ \aalpha \in \Nd \st \len{\aalpha} = l \}.
\]

Surprisingly, there are several ways to construct the slices of degree~$l$, by
induction on either the right or left and by the order of the sub-calls,
\[
  \calC^d_l
  = \biguplus_{i=0}^l \left\{
    (i, \cbbeta) \in \Nd \st \cbbeta \in \calC^{d-1}_{l-i} \right\}
  = \biguplus_{i=0}^l \left\{
    (\tbbeta, i) \in \Nd \st \tbbeta \in \calC^{d-1}_{l-i} \right\}.
\]
The first equality makes the family ordered by lex and the second one by colex,
therefore making~$\calAdk$ ordered by grlex or grcolex.

In order to have monomials ordered by decreasing exponents on the successive
variables~$X_0$,\dots\linebreak[0] $X_{d-1}$, we choose
\[
  \calC^d_l = \biguplus_{i=0}^l \left\{
    (l - i, \cbbeta) \in \Nd \st \cbbeta \in \calC^{d-1}_i \right\},
\]
which orders~$\calC^d_l$ by symlex.
And so we prove that~$\calAdk$ is ordered by (\coqe{grsymlex lt}).

\subsection{Conclusion}

We have presented a {\Rocq} comprehensive formalization of many properties on
orders, especially on monomial orders.
The initial motivation was the ordering of multi-indices (on~$\Nd$) for the
formalization of multivariate polynomials, but we end with a comprehensive
formalization of properties and lemmas about generic orders.
Even if some part is really well-known (such as transitivity or irreflexivity),
we focus on proof engineering, we then bring up some subtleties about
definitions, and we end up with a very usable and comprehensive library of
properties and how they are linked, that amounts to more than 50 definitions
and nearly 700 lemmas.
Moreover, the definition of graded orders as a high-level operator has led us
to elegant and factorized proofs.

Thanks to our well-chosen definitions, there is no particularly difficult
proof.
Note also that it is not easy to have a reference of all these statements,
especially as formal lemmas must encompass {\unintuitive} cases, for instance
{\nontotal} orders or what happens on empty vectors.
A difficulty is the handling in a single setting of both strict and
{\nonstrict} orders, especially in degenerate cases.

\subsection{Weighted Orders}

The main perspective of this work are the weighted orders~\cite{clo:iva:15}.
We assume we have a ring~$\K$ (and not only a monoid as before) so that we have
a dot product denoted by ``$\cdot$''.
We also assume that the strict order~$\ltK$ is compatible not only with~$+$,
but with $\times \ell$ for positive $\ell\in\K$.
Then, there is a generalization of orders on~$\Kd$: a weighted order is
characterized by an order~$\ltK$ on~$\K$ and a matrix~$W$ of type
$\K^{d,m}$.
Then if we denote the blocks of the matrix as $W=[\ww_0\ |\ W_{1\dots m-1}]$,
where $\ww_0\in \Kd$ and $W_{1\dots m-1}$ is in $\K^{d,m-1}$,
\begin{equation}
  \label{e:wo}
  \xx <^W \yy \EQUIVDEF
  \left\{
  \begin{array}{l}
    \xx \cdot \ww_0 \ltK \yy \cdot \ww_0, \mbox { or}\\
    \xx \cdot \ww_0 = \yy \cdot \ww_0 \CONJ \xx <^{W_{1\dots m-1}} \yy.
  \end{array}
  \right.
\end{equation}
This matrix vision encompasses all the defined orders (lex, grsymlex, and so
on) and also provides generic lemmas.
For instance for a square matrix on~$\R$, the order is total if and only if the
matrix is invertible.
The graded operator corresponds to creating a block matrix with the first
column filled with ones and the initial matrix.

The fact that the matrix may be rectangular adds to the generality of the
definition, but also to the redundancy, as several matrices correspond to the
same order.
An interesting point is that we can relate some matrices to the alternative
definitions: for instance, for grsymlex, three matrices correspond
to~\eqref{e:grsymlex}, \eqref{e:grsymlex-equiv-1}, and
\eqref{e:grsymlex-equiv-2} (the first and last matrices being square).

More generally, for an order with a real invertible square matrix~$W$, if we
right-multiply~$W$ by a square upper-triangular matrix with positive diagonal
terms, then the corresponding order is the same.
Thus, using a~$QR$ decomposition of a matrix of an order~$W$ ($=QR$), it is
therefore possible to take~$Q$ as a canonical form of the order.

Even if appealing, our formalization attempts have not been that successful.
In particular, mathematics mostly consider square matrices for~$W$, which is
easier for invertibility, equivalence, and so on.
But it makes it difficult to use the induction underlying~\eqref{e:wo}.
There is work left to precisely understand in which cases the matrix should be
rectangle or square and have a usable formalization of weighted orders.

\bibliographystyle{plainnat}
\bibliography{biblio}

\end{document}